\documentclass[aps,showpacs,floats]{revtex4}
\usepackage{amssymb}
\usepackage{graphicx}
\usepackage{bm}

\begin{document}

\bibliographystyle{apsrev}

\def\v#1{{\bf #1}}
\newcommand{\la}{\langle}
\newcommand{\ra}{\rangle}
\newcommand{\lc}{\lowercase}

\newcommand{\hv}{\hat{v}}

\newcommand{\nn}{\nonumber}

\newcommand{\be}{\begin{equation}}
\newcommand{\ee}{\end{equation}}
\newcommand{\bea}{\begin{eqnarray}}
\newcommand{\eea}{\end{eqnarray}}

\newcommand{\tphi}{\tilde{\varphi}}
\newcommand{\tthe}{\tilde{\vartheta}}
\newcommand{\trho}{\tilde{\rho}}
\newcommand{\tE}{\tilde{E}}

\newcommand{\Q}{(0,0,$\frac{\pi}{c}$)}
\newcommand{\Qh}{(0,0,$\frac{\pi}{2c})$}

\newcommand{\UPD}{UP\lc{d}$_2$A\lc{l}$_3$}
\newcommand{\LOS}{L\lc{a}O\lc{s}$_4$S\lc{b}$_{12}$}
\newcommand{\POS}{P\lc{r}O\lc{s}$_4$S\lc{b}$_{12}$}
\newcommand{\POSX}{P\lc{r}$_{1-x}$L\lc{a}$_x$O\lc{s}$_4$S\lc{b}$_{12}$}

\title{Field dependent mass enhancement in \POSX~ from aspherical Coulomb scattering}

\author{Gertrud Zwicknagl$^{(1)}$, Peter Thalmeier$^{(2)}$ and Peter Fulde$^{(3,4)}$}
\affiliation{
$^{(1)}$Technische Universit{\"a}t Braunschweig, 38106 Braunschweig, Germany\\
$^{(2)}$Max-Planck-Institut f{\"u}r Chemische Physik fester Stoffe, 01187 Dresden, Germany\\
$^{(3)}$Max-Planck-Institut f{\"u}r Physik komplexer Systeme, 01187 Dresden, Germany\\
$^{(4)}$Asia Pacific Center for Theoretical Physics, Pohang, Korea}

\begin{abstract}
The scattering of conduction electrons by crystalline electric field (CEF) excitations may enhance their effective quasiparticle mass similar to scattering from phonons. A wellknown example is Pr metal where the isotropic exchange scattering from inelastic singlet-singlet excitations causes the mass enhancement. An analogous mechanism may be at work in the skutterudite compounds \POSX~ where close to x=1 the compound develops heavy quasiparticles with a large specific heat $\gamma$. There the low lying CEF states are singlet ground state and a triplet at $\Delta$ = 8 K. Due to the tetrahedral CEF the main scattering mechanism must be the aspherical Coulomb scattering. We derive the expression for mass enhancement in this model including also the case of dispersive excitations. We show that for small to moderate dispersion there is a strongly field dependent mass enhancement due to the field induced triplet splitting. It is suggested that this effect may be seen in \POSX~ with suitably large x when the dispersion is small.
\end{abstract}


\maketitle

\section{Introduction}

\label{sect:introduction}

The filled-skutterudite compound PrOs$_4$Sb$_{12}$ has recently obtained
considerable attention. There are several reasons for that. It is a heavy
fermion ($\gamma\sim 350-500 mJ/molK^2$) superconductor
with a transition temperature of $T_c$(Pr) = 1.85
K. This temperature is larger than the one of the related system
LaOs$_4$Sb$_{12}$ which is $T_c$(La) = 0.74 K. A number of experiments,
like those on Sb-NMR relaxation rate in \POSX~ \cite{Yogi06} suggest
that the superconducting order parameter is of the conventional isotropic
$s$-wave type with possible admixture of higher harmonics depending on the
Pr content \cite{Parker08}. However, questions and ambiguities remain.
They concern the experimental findings
for the penetration depth \cite{Chia03,MacLaughlin08} and initial studies of the thermal
conductivity in a rotating magnetic field \cite{Izawa03}. For example, the former
suggest a possible nodal
structure while the latter in addition  points towards two distinct superconducting
phases. The observed enhancement of the superconducting transition temperature
by more than a factor of two when La is replaced by Pr seems surprising at
first sight. It is well known that when Pr ions are added as impurities to an
$s$ wave superconductor like LaPb$_3$ it suppresses the superconducting
transition temperature rather efficiently. So why does it enhance $T_c$ in the
present case? Since the phonons in LaOs$_4$Sb$_{12}$ and PrOs$_4$Sb$_{12}$ are
very nearly the same, the enhancement must come from the two 4$f$ electrons
which Pr$^{3+}$ has. The heavy fermion behavior of PrOs$_4$Sb$_{12}$ seems
puzzling too. It shows up in a large specific heat jump $\Delta C/T_c \simeq$
500 mJ/(mol K$^2$) at $T_c$ and also in a large effective mass in de Haas-van
Alphen experiments. The Kondo effect cannot be the origin of the heavy
quasiparticles since the 4$f^2$ electrons are well localized with a Hund's rule
total angular momentum $J = 4$ and a non-Kramers ground state.
The key to the enhancement of $T_c$ and the
formation of heavy quasiparticle excitations lies in the crystalline electric
field (CEF) splitting of the  $J = 4$ multiplet  (Sect.~\ref{sect:CEF}),
together with the aspherical Coulomb scattering mechanism of conduction electrons from 
CEF excitations \cite{Goremychkin04,Chang07}.
The  self energy and effective mass enhancement due to this mechanism will be calculated
in Sect.~\ref{sect:mass}. The CEF states, their excitation energies and matrix elements are
modified by an external field. The ensuing effective mass dependence on the field
which is the main topic of the present work is calculated in Sect.~\ref{sect:field} for dispersionless excitations and
in Sect.~\ref{sect:disp} for the case with dispersive quadrupolar excitons . Some numerical
results are discussed in  Sect.~\ref{sect:discussion} and Sect.~\ref{sect:conclusion} finally gives the conclusions.

%
%%%%%%%%%%%%%%%%%%%%%%%%%%%%%%%%%%%%%%%%%%%%%%%%%%%%%%%%%%%%%
\begin{figure}
\raisebox{-1.5cm}
{\includegraphics[width=8.0cm]{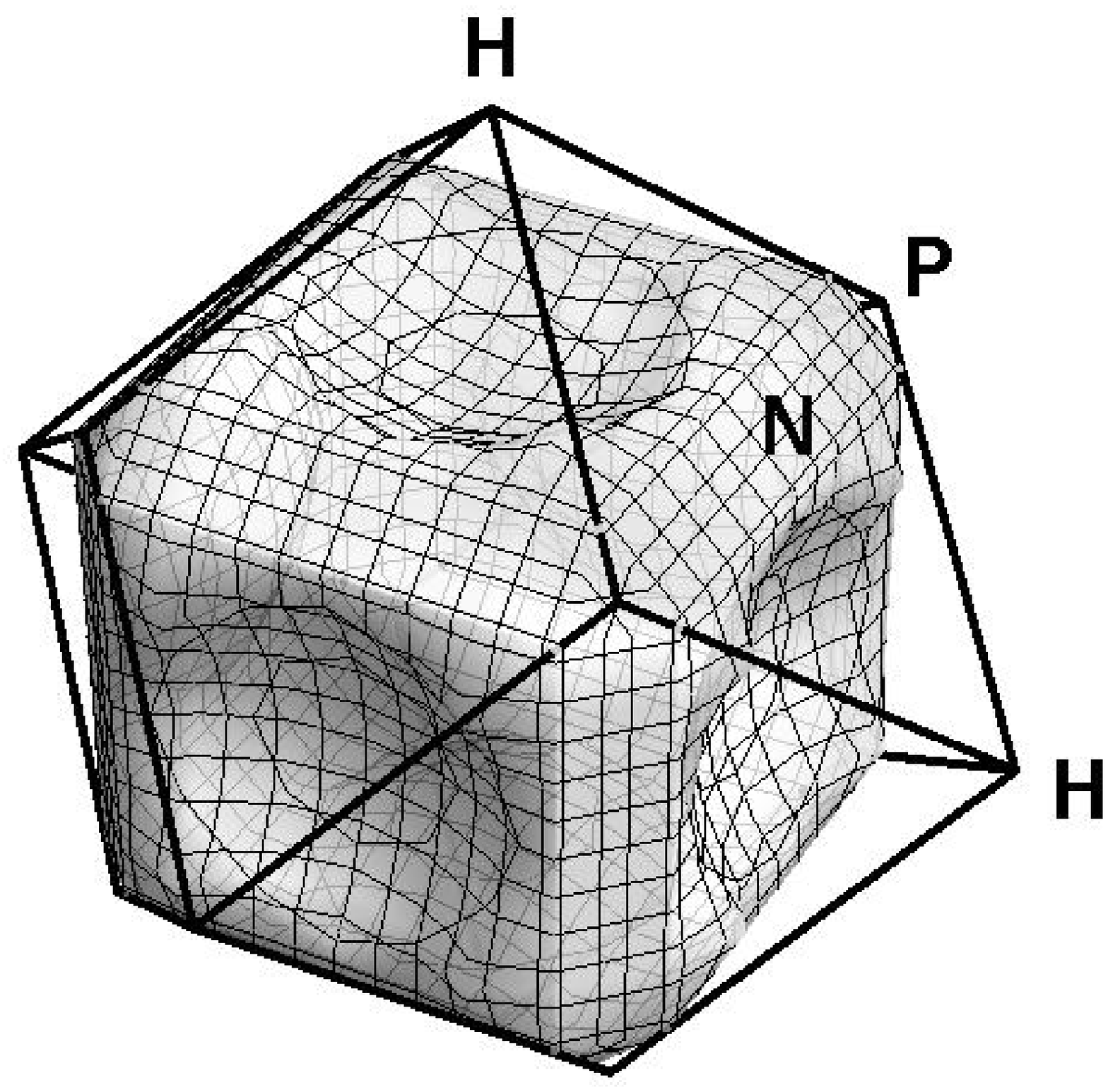}}
\includegraphics[width=8.0cm]{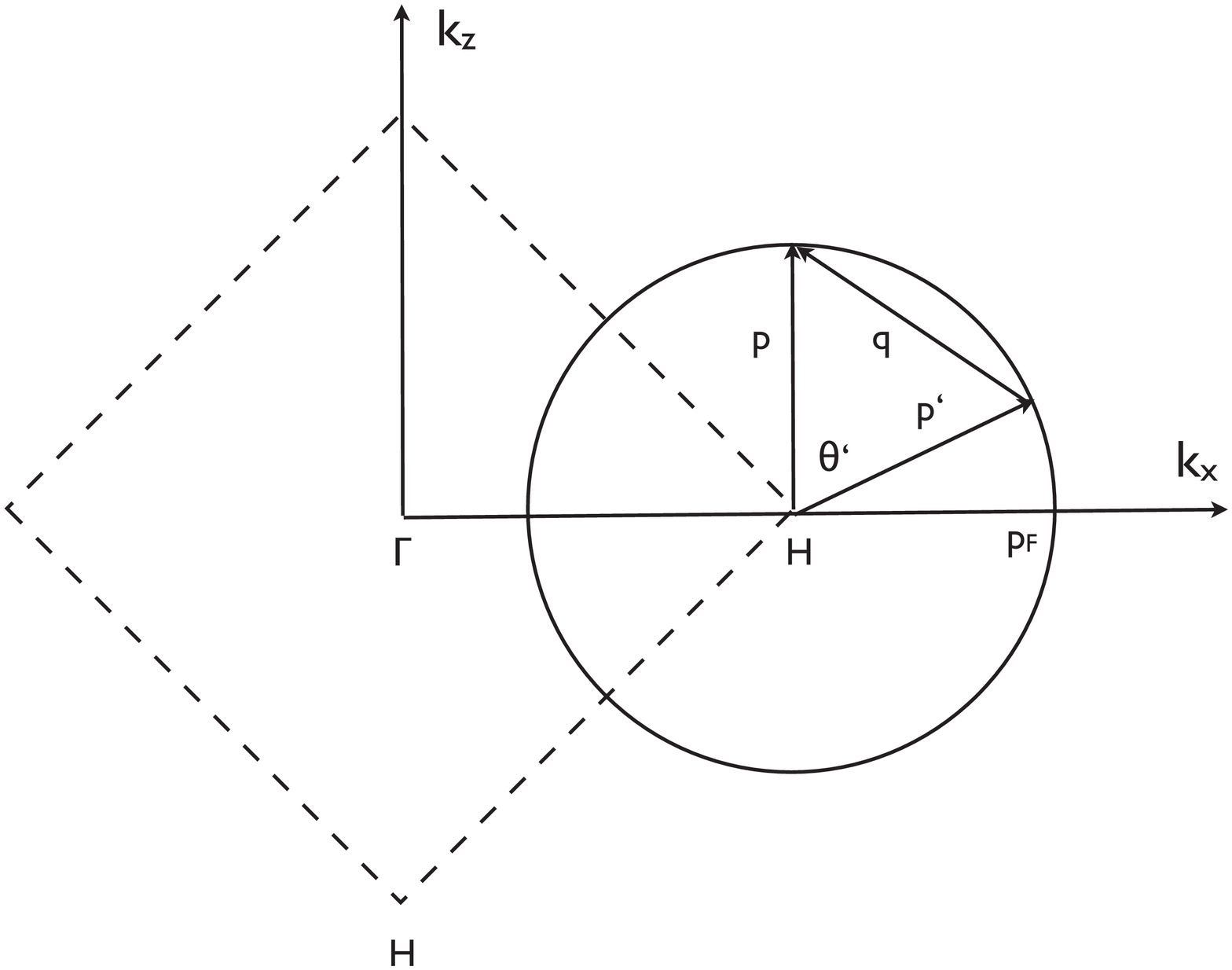}
\caption{Left (a): Fermi surface of n.n.n. tight binding model in {\em hole} representation in the bcc Brillouin zone.
Right (b): schematic Fermi surface in {\em electron} representation in the 2D projected Brillouin zone.
 It consists of spheroids around the equivalent H-points $(\frac{2\pi}{a},0,0)$. The polar angle $\theta$ of \v q is given by
 $\theta = \frac{1}{2}(\pi-\theta')$. Furthermore we have $q=2p_F\sin\frac{\theta'}{2}=2p_F\cos\theta$ where $p_F$ is the
 Fermi momentum. The geometric restrictions
 require $0\leq q \leq 2p_F$ and  $0\leq\theta\leq\frac{\pi}{2}$.}
\label{fig:Fig0}
\end{figure}
%%%%%%%%%%%%%%%%%%%%%%%%%%%%%%%%%%%%%%%%%%%%%%%%%%%%%%%%%%%%%
%

\section{The CEF states of P\lc{r} in T$_h$ symmetry and their interactions}

\label{sect:CEF}

From inelastic
neutron scattering the CEF energy levels are known. The compound has tetrahedral $T_h$ site
symmetry for Pr. The data are explained best by a CEF ground state

%1
\begin{equation}
\mid \Gamma_1 \rangle = \frac{\sqrt{30}}{12} \left( \mid +4 \rangle + \mid -4
\rangle \right) + \frac{\sqrt{21}}{6} \mid 0 \rangle
\label{1.01}
\end{equation}

\noindent with a low-lying triplet excited state at an energy of $\Delta = 8 K$
\cite{Goremychkin04,Kuwahara04,Shiina04a,Shiina04b}. The other CEF levels are so high
in energy that they can be
neglected. The $\Gamma_t$ triplet state of $T_h$ symmetry is a superposition
of two triplets $\Gamma_4$ and $\Gamma_5$ of O$_h$ symmetry. More specifically
one finds \cite{Shiina04a,Shiina04b}

%2
\begin{equation}
\mid \Gamma_t, m \rangle = \sqrt{1-d^2} \mid \Gamma_5, m \rangle + d \mid
\Gamma_4, m \rangle \qquad , ~~ m = 1 ... 3
\label{1.02}
\end{equation}

\noindent with states of O$_h$ symmetry given by

%4
\begin{eqnarray}
\mid \Gamma_5, \pm \rangle & =&  \pm \sqrt{\frac{7}{8}} \mid \pm 3 \rangle \mp
\sqrt{\frac{1}{8}} \mid \mp 1 \rangle \;\qquad
\mid \Gamma_5, 0 \rangle  =  \sqrt{\frac{1}{2}} \left( \mid +2 \rangle - \mid
-2 \rangle \right) \nonumber\\
\mid \Gamma_4, \pm \rangle & = & \mp \sqrt{\frac{1}{8}} \mid \mp 3 \rangle \mp
\sqrt{\frac{7}{8}} \mid \pm 1 \rangle \;\qquad
\mid \Gamma_4, 0 \rangle  =  \sqrt{\frac{1}{2}} \left( \mid +4 \rangle - \mid
-4 \rangle \right) ~~~.
\label{1.04}
\end{eqnarray}
%\newpage

The conduction electrons interact with the CEF energy levels of the Pr$^{3+}$
ions. The most important ones are the isotropic exchange interactions and the
aspherical Coulomb scattering. The former is of the form

%5
\begin{equation}
H_{\rm ex}(i) = -2 \left( g_J - 1 \right) J_{\rm ex} \sum_{{\bf k}{\bf
	q}\sigma\sigma'}( {\bf s}_{\sigma\sigma'}\cdot{\bf J}_i) c^\dagger_{{\bf k-q}\sigma'} c_{{\bf
	k}\sigma}
\label{1.05}
\end{equation}

\noindent where $c^\dagger_{{\bf k}\sigma} (c_{{\bf k}\sigma})$ are the creation
(annihilation) operators for conduction electron with momentum ${\bf k}$ and
spin $\sigma$ while ${\bf s}$ is their spin operator. Furthermore $g_J$ is the
Land\'e factor. The aspherical Coulomb  interaction  in local orbital basis is given by
\cite{Fulde70}

%6
\begin{equation}
H_{\rm AC}(i) = \left( \frac{5}{4\pi}\right)^{\frac{1}{2}} \sum_{kk'\sigma}
\ \sum^{+2}_{m=-2} I_2 \left( k's;kd \right) Q_2 \left[ Y^m_2 ({\bf J}_i) c^\dagger_{k's
	\sigma} c_{kdm \sigma} + h.c. \right]~~~.
\label{1.06}
\end{equation}

\noindent Here $c_{klm \sigma}$ destroys a conduction electron with momentum
  $k=|{\bf k}|$, in a $l = 2$ state with azimuthal quantum number $m$ and spin
  $\sigma$ and $c^\dagger_{k's \sigma}$ creates one with momentum $k'$ in a
  $l = 0$ state. The Coulomb integral $I_2$ is defined, e.g., in
  Ref.~\onlinecite{Fulde70} and $Q_2$ is the quadrupole moment of the Pr$^{3+}$
  ion. The operators $Y^m_2({\bf J})$ are given by

%7
\begin{eqnarray}
Y^0_2 & = & \left( 2/3 \right)^{1/2} \left[ 3 J^2_z - J (J+1)
  \right] / N_J \nonumber \\
Y^\pm_2 & = & \pm \left( J_z J^\pm + J^\pm J_z \right) / N_J \nonumber \\
Y^{\pm 2}_2 & = & \left( J^\pm \right)^2 / N_J
\label{1.07}
\end{eqnarray}

\noindent with $N_J = (2/3)^{1/2}(2J^2-J)$. $H_{\rm AC}$ leads to a transfer of
angular momentum $l = 2$ between the conduction electrons and the incomplete
4$f$ shell. It is a quadrupolar type of interaction.

An important feature of PrOs$_4$Sb$_{12}$ is the experimental finding that the
low-energy triplet state has a small value of $|d| = 0.26$ with the implication
that the inelastic scattering of the conduction electrons is predominantly of
quadrupolar character. With this information the two features pointed out
above, i.e., the increase of $T_c$ when La is replaced by Pr and the heavy
quasiparticle mass can be understood quantitatively \cite{Chang07}. As has been known for a long
time quadrupolar inelastic scattering of conduction electrons by low-energy CEF
levels enhances Cooper pairing since these excitations act similarly as a
localized phonon mode. The difference is that phonons are related to changes in
the ion position while intra-atomic quadrupolar CEF excitations are related to
changes of the 4$f^2$ wavefunction.

Also the heavy quasiparticle mass is related to the inelastic scattering
processes of the conduction electrons. This feature has been previously
exploited to explain the mass enhancement found in Pr metal  using the
isotropic dipolar exchange interaction H$_{ex}$ \cite{Fulde83}.
As mentioned above in \POSX~ the aspherical Coulomb scattering
H$_{ac}$ is dominant over exchange scattering. For this model a
quantitative calculation of the changes in $T_c$ and the mass enhancement
as function of Pr concentration are found in Ref.~\onlinecite{Chang07}.

The aim of the present communication is to extend the previous calculations by
including the effect of an external magnetic field on the mass enhancement. A
field splits the triplet states and leads to a decrease of the excitation
energy of one of the three states, at least for small tetrahedral admixture d as in \POS.
Therefore an increase of the effective mass with increasing field is expected in this case.

The situation is different for Pr metal mentioned before  where ground state and first excited
state are two singlets. There a magnetic field pushes the two energy levels
apart and hence increases the excitation energy. As a consequence the effective
quasiparticle mass decreases with increasing external field in agreement with experimental
findings\cite{Forgan81}. In the present singlet-triplet model this case would be realised for
$|d|>0.65$.

%
%%%%%%%%%%%%%%%%%%%%%%%%%%%%%%%%%%%%%%%%%%%%%%%%%%%%%%%%%%%%%
\begin{figure}
\includegraphics[width=8.0cm]{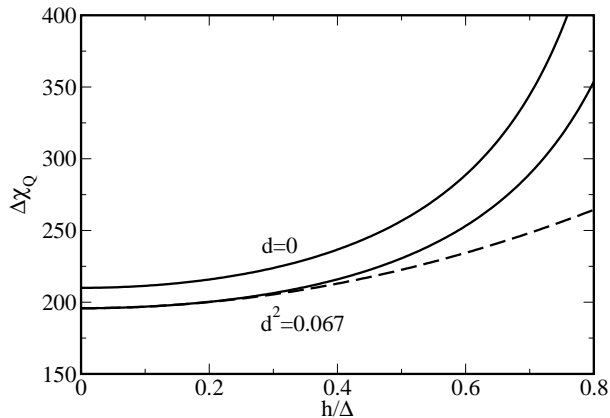}
\caption{Static (normalised) quadrupolar susceptibility $\Delta\chi_Q$  as function of magnetic field with ($d^2=0.067$ or $ |d|=0.26$)  and without ($d^2=0$) tetrahedral CEF. Full lines correspond to evaluation with Eq.~(\ref{2.24}) while the dashed line corresponds to the low field expansion Eq.~(\ref{2.25}).  Approaching the $\Gamma_s-\Gamma^+_t$ level crossing leads to an increasing $\chi_Q$  and mass enhancement which becomes singular at the crossing $h_c/\Delta = (1-\delta^2)^{-1}$. The latter is pushed to higher field for increasing $d^2$ which reduces the increase of $\chi_Q$. For $d^2\geq 0.42$ $\chi_Q$ {\em decreases} with field strength because the tetrahedral CEF leads to $\Gamma_s-\Gamma^0_t$  repulsion. The mass enhancement is proportional to the quadrupolar susceptibility with $\delta m^*/m = g_{eff}(\Delta\chi_Q)$ and $g_{eff}=(\tilde{g}/\Delta)\bar{f}$. For $d^2=0.067$ and  $g_{eff} = 0.077$ (Sect.~\ref{sect:disp}) one has $(\delta m^*/m)_{h=0} =16$.}
\label{fig:Fig1}
\end{figure}
%%%%%%%%%%%%%%%%%%%%%%%%%%%%%%%%%%%%%%%%%%%%%%%%%%%%%%%%%%%%%
%

\section{Self energy and Mass renormalisation}

\label{sect:mass}

We start out by specifying the electronic part of the Hamiltonian for the system
La$_{1-x}$Pr$_x$Os$_4$Sb$_{12}$. It is of the type

%8
\begin{equation}
H = H_{\rm el} + H_{\rm CEF} + H_{\rm AC} + H_{\rm ex} + H_{\rm Z}~~~.
\label{2.08}
\end{equation}

\noindent Here  H$_{\rm el}$  is of the conventional form
and need not be explicitely written down. It contains the conduction band dispersion
which may be described by a n.n.n. tight binding model \cite{Westerkamp} according to

\begin{equation}
\epsilon_{\bf k\sigma}=t\cos\frac{1}{2}k_x\cos\frac{1}{2}k_y\cos\frac{1}{2}k_z
+t'(\cos k_x+\cos k_y+\cos k_z)
\end{equation}

with t=174 meV and t'=-27.84 meV. The transfer integrals t and t' are chosen so
as to reproduce the observed linear specific heat coefficient $\gamma = 36 \textrm{mJ/(mole K}^2)$ of the
non-f reference compound LaOs$_{12}$Sb$_{12}$ \cite{Bauer01}.
In the {\em electron} picture the associated Fermi surface
consists of H-centered spheroids with a Fermi radius $p_F\simeq 0.7(2\pi/a)$, see Fig.~\ref{fig:Fig0}b. Aside from
subtle effects this is quite similar to the LDA FS in Ref.~\onlinecite{Miyake03} (note that in this reference
the FS is depicted in the {\em hole} picture as in  Fig.~\ref{fig:Fig0}a). It corresponds to a single band 
originating in Sb-4p states.

The CEF and Zeeman Hamiltonian are

%9
\begin{equation}
H_{\rm CEF}+H_Z = \sum_{i,\Gamma n}E_\Gamma \mid \Gamma_n (i) \rangle \langle \Gamma_n (i) \mid +
g_J \mu_B \sum_i {\bf J} (i)\cdot {\bf H}~~~.
\label{2.09}
\end{equation}

\noindent
\noindent The external magnetic field is denoted by ${\bf H}$, $g_J$ is the
Land\'e factor and $\mu_B$ is Bohr's magneton. Furthermore $i$ labels the Pr sites, and
$|\Gamma_n\rangle$ denotes the singlet ground-state $| \Gamma_s \rangle$ and the triplet
$| \Gamma^n_t \rangle$ (see Eq.~(\ref{1.02})) with energies $E_s = 0$ and $E_t =$ 8 K,
respectively. We assume that not
only the phonons but also their local interactions with the electrons are independent
of partial replacements of La by Pr.

%The local interaction Hamiltonians
%corresponding to $H_{\rm el-CEF}$ have been written down in (\ref{1.05}) and
%(\ref{1.06}) for a spatially isotropic situation.

As pointed out before the system has $T_h$ symmetry but since the CEF
transition can be reduced to those between states of cubic symmetry (see
(\ref{1.02},\ref{1.04})) we specialize (\ref{1.06}) to cubic
symmetry. In that case the aspherical Coulomb  interaction written in a basis of Bloch states becomes

%10
\begin{equation}
H_{\rm AC} (i) = g\sum_{{\bf kq}\sigma} \sum_{\alpha
  \beta\;\; cycl. \atop \sigma} O^i_{\alpha \beta} \hat{q}_\alpha \hat{q}_\beta  c^\dagger_{{\bf k-q} \sigma}
  c_{{\bf k}\sigma}e^{i{\bf k}{\bf R}_i}
\label{2.10}
\end{equation}

\noindent
with $O_{\alpha \beta} = \frac{\sqrt{3}}{2}(J_\alpha J_\beta + J_\beta J_\alpha)$, $\hat{q}_\alpha  =
q_\alpha/|{\bf q}|$ and $\alpha\beta = yz, zx, xy$ denoting the three quadrupole operators with $\Gamma_5$ symmetry.
The remaining $\Gamma_3$ quadrupole terms are neglected since they do not couple to the singlet-triplet excitations.
The coupling constant g may in principle be determined by experiments. A derivation of (\ref{2.10}) may be obtained 
from Ref.~\onlinecite{Teitelbaum76}.

In order to determine the effective mass $m^*$ of the quasiparticles at zero
temperature one must calculate the Green's function of conduction electrons.

%13
\begin{equation}
G \left( {\bf p}, \omega \right) = \frac{1}{\omega - \epsilon ({\bf p}) -
\Sigma({\bf p}, \omega)}~~~.
\label{2.13}
\end{equation}

\noindent The effective mass enhancement due to interactions of the conduction
electrons follows from

%14
\begin{equation}
\frac{m^*}{m} = 1 - \left. \frac{\partial \Sigma(p_F,\omega)}{\partial \omega}
\right|_{\omega = 0}
\label{2.14}
\end{equation}

where $p_F$ is the Fermi momentum and $m$ is the reference quasiparticle mass including band effects and
 effects of electron-phonon coupling. Neglecting vertex corrections the irreducible electron self
energy  $\Sigma ({\bf p}, \omega)$  due to H$_{\rm AC}$ is given by

%15
\begin{equation}
\Sigma \left( {\bf p}, \omega \right) = g^2\sum_{\alpha\beta,n} \int \frac{d^3 q}{(2 \pi)^3}
|\Lambda^n_{\alpha\beta}(\hat{\bf q})|^2
 \int\frac{d \omega'}{2 \pi} D_n\left( {\bf q}, \omega \right) G \left( {\bf p} -
	 {\bf q}, \omega - \omega' \right) ~~~.
\label{2.15}
\end{equation}

\noindent
Here $D_n({\bf k}, \omega)$ denotes the boson propagator of CEF excitations.
It is related to the dynamical quadrupolar susceptibility of the CEF system. In the present case we
will neglect effective RKKY type interactions between CEF states on different
sites therefore the boson propagator is local ({\bf q}- independent).
The momentum dependence of the bare vertex $\Lambda^n_{\alpha\beta}(\hat{\bf q})$
is due to the quadrupolar $\Gamma_5$ form factors in Eq.~(\ref{2.10}). It is defined as

%16
\begin{equation}
\Lambda_{\alpha\beta}^n(\hat{\bf q})=g\hat{q}_\alpha\hat{q}_\beta
 \langle \Gamma_s \left| O_{\alpha\beta} \right| \Gamma^n_t
  \rangle  ~~~.
\label{2.16}
\end{equation}

The self energy due to exchange scattering which involves the magnetic
susceptibility can be safely neglected because of the smallness of $d^2$ (see
(\ref{1.02})) as a more detailed investigation including matrix elements and
coupling constants shows. The  propagator of the local singlet-triplet boson
excitations is given by

%17
\begin{equation}
D_n \left( {\bf q}, \omega \right) = D_n(\omega)= \frac{2\delta_n}{\delta_n^2 -
  \omega^2} ~~~.
\label{2.17}
\end{equation}

\noindent
Here the field dependent singlet-triplet excitation energies are given by
$\delta_n(H)=\epsilon_t^n(H)-\epsilon_s(H)$ (n = +,0,-).

The self-energy in Eq.~(\ref{2.15}) can be evaluated following Migdal's integration procedure (see. e. g. \cite{ScalapinoParks}).
This method exploits the fact that the summation over fermionic states in the vicinity of the Fermi surface separates into independent summations over energy and degeneracy variables. Since the relevant excitation energies are of the order of the CEF excitation $\delta_n(H) $ the dominant contribution to the integral in Eq.~(\ref{2.15}) comes from electronic states with $\left| \epsilon ({\bf p}) \right| \sim \delta_n(0) \ll W$.
First one replaces the integral over {\bf q} by an equivalent integration over  ${\bf p}' = {\bf p} - {\bf q}$
where the external momentum ${\bf p}$ is kept fixed.
Then one restricts the frequency integration to a shell $|\omega'|\ll \epsilon_c$ around the Fermi surface such that
$\delta_\alpha\ll\epsilon_c \ll W$ is fulfilled. Here 2W is the conduction band width. In this shell we may approximate the momentum space integral by

 %18
\begin{equation}
\int \frac{d^3 p'}{(2 \pi)^3} =
\frac{N(0)}{4\pi}\int_0^{2\pi} d\phi' \int_0^\pi \sin\theta' d\theta' \int_{-\epsilon_c}^{\epsilon_c} d\epsilon'
%\equiv N(0) \frac{1}{2p^2_F} \frac{4}{15}\int^{2p_F}_0 dqq \int_{-\epsilon_c}^{\epsilon_c} d \epsilon'
~~~.
\label{2.18}
\end{equation}

\noindent
where $\theta', \phi'$ are the polar and azimuthal angles of $\hat{\bf p}'$. Furthermore $N(0)$ is the conduction electron density of states  per spin at the Fermi energy ($\epsilon_F=0$).
Using the analytical properties of the self-energy $\Sigma(\omega)$ which imply that
$ \textrm{sign Im}\, \Sigma(\omega)=-\textrm{sign}\, \omega$  the $\epsilon '$-integration gives for $\delta_n(H) $
\begin{equation}
\Sigma(\omega)=  g^2 N(0) \sum_{\alpha \beta, n} \langle q_\alpha ^2 q_\beta ^2\rangle
|\langle \Gamma_s |O_{\alpha \beta} | \Gamma_t ^n\rangle|^2
\int \frac{d\omega'}{2\pi} \frac{2\delta_n}{(\omega-\omega')^2-\delta_n^2+i\eta}\,\textrm{sign}(\omega')
2 \arctan\frac{\epsilon_c}{|\textrm{Im} \Sigma(\omega')|} \quad.
\label{2.18a}
\end{equation}

We solve the self-consistency equation for the imaginary part $\textrm{Im}\, \Sigma(\omega)$
\begin{equation}
\textrm{Im}\, \Sigma(\omega)=-g^2 N(0) \sum_{\alpha \beta, n} \langle q_\alpha ^2 q_\beta ^2\rangle
|\langle \Gamma_s |O_{\alpha \beta} | \Gamma_t ^n\rangle|^2
\sum_{\rho=\pm 1} \textrm{sign}(\omega+\rho \delta_n(H)) \arctan\frac{\epsilon_c}{|\textrm{Im}\, \Sigma(\omega+\rho \delta_n(H))|}
\end{equation}
from which we subsequently deduce the real part by Kramers-Kronig transformation. The explicit form immediately shows that
\begin{equation}
|\textrm{Im}\, \Sigma(\omega)| \leq \pi g^2 N(0)  \sum_{\alpha \beta, n} \langle q_\alpha ^2 q_\beta ^2\rangle
|\langle \Gamma_s |O_{\alpha \beta} | \Gamma_t ^n\rangle|^2 =\frac{\hbar}{\tau}\quad .
\end{equation}
It is important to note that $\textrm{Im} \Sigma(\omega)$ exhibits discontinuous jumps at the energies corresponding to the
singlet-triplet excitations. This feature is a direct consequence of the assumption that the CEF excitations are long-lived and dispersionless bosonic excitations. Of particular interest is the discontinuity at $\delta_+(H)$
\begin{equation}
|\textrm{Im}\, \Sigma(\delta_+(H)+\eta)-\textrm{Im}\, \Sigma(\delta_+(H)+\eta)| \geq
\frac{\hbar}{\tau} \frac{2}{\pi}
 \arctan\frac{\epsilon_c \tau}{\hbar} \quad .
\end{equation}
This discontinuity in the imaginary part inevitably implies a logarithmic singularity in the real part $\textrm{Re}\, \Sigma(\omega)$ which, in turn, leads to an unphysical divergence in the effective mass for $\delta_+(H)\to 0$.

In the limit $\epsilon_c \to \infty$ where
$2 \arctan\frac{\epsilon_c}{|\textrm{Im} \Sigma(\omega')|} \to \pi$ the result agrees with that of non-selfconsistent
second order perturbation theory.
In this case differentiating the self energy  with respect to  $\omega$ under the integral and using integration by parts  one finally gets from Eq.~(\ref{2.14}):

%19
\begin{eqnarray}
%\frac{m^*}{m} &=& 1 + g^2 N(0) \sum_{\alpha\beta,n}\langle\hat{q}^2_\alpha \hat{q}^2_\beta\rangle
%\frac{2|\langle \Gamma_s \left| O_{\alpha\beta} \right| \Gamma^n_t\rangle|^2}{\delta_n}\quad\mbox{or}
%\nonumber\\
\frac{m^*}{m}& =& 1 + g^2 N(0)\bar{f}\chi_Q(H);\quad
%\chi_Q(H)= \sum_{\alpha\beta}\chi^{\alpha\beta}_Q(H)
\nonumber\\
\chi_Q (H) &=& \sum_{\alpha\beta,n} \frac{2\left| \langle
\Gamma_s (H) \left| O_{\alpha\beta} \right| \Gamma^n_t (H) \rangle\right|^2}{\delta_n(H)}~~~.
\label{2.19}
\end{eqnarray}

The directional average (with respect to polar and azimuthal angles  $\theta,\phi$ of $\hat{\bf q}$)  for quadrupolar form factors $\bar{f}=\langle\hat{q}^2_\alpha \hat{q}^2_\beta\rangle$ = $\frac{1}{15}$ is a constant.  Furthermore $\chi_Q(H)$ in Eq.~(\ref{2.19}) is  the field-dependent static uniform quadrupolar susceptibility. Note that the form factor average can be trivially factored out as a constant (1/15) only because in the present local approximation the boson propagator is momentum independent, i.e. the CEF excitations are dispersionless. The more general case will be discussed below.

\section{Field dependence of the effective mass: dispersionless model}

\label{sect:field}

When a magnetic field is applied to the sample the field dependence of the effective mass is completely determined by that of the quadrupolar susceptibility in Eq.~(\ref{2.19}). To calculate this quantity one first has to know the singlet-triplet excitation energies  $\delta_n (H)$  and the eigenstates and matrix elements in applied field. They were given by Shiina {\it et al} \cite{Shiina04a,Shiina04b} in closed form for field applied along cubic symmetry directions. We use these results  in the following. The CEF eigenstates are denoted as
singlet  $|\Gamma_s\rangle$ and  triplet $|\Gamma^n_t\rangle$ (n = +,0,-), respectively. The CEF and Zeeman Hamiltonian can be easily mapped to a pseudospin basis \cite{Shiina04a,Shiina04b} and then diagonalised. In pseudospin basis the zero-field singlet is denoted by $|0,0\rangle$ and the triplet by $|1,m\rangle$ (m=1,0,-1). For field ${\bf H} \parallel [001]$ the field-split CEF levels and mixed eigenstates are then given in Table~\ref{table:table1}. The field dependence of $\delta_n(H)$ has recently been determined by INS eperiments \cite{Raymond08}.

The mixing coefficients $u,v$ are determined by the matrix elements of the dipolar operator {\bf J} in the Zeeman term which may be expressed by $\alpha = 5/2-2d^2, \beta =2\sqrt{5/3}d, \delta=\beta/\alpha$. They are given by

%22
\begin{eqnarray}
v&=&-sgn(y)[\frac{1}{2}\bigl(1-\frac{\Delta}{\tilde{\Delta}}\bigr)]^\frac{1}{2};\qquad u=(1-v^2)^\frac{1}{2}\nonumber\\
\tilde{\Delta}&=&[\Delta^2+4\delta^2h^2]^\frac{1}{2};\qquad h\equiv g\mu_B\alpha H
\label{2.22}
\end{eqnarray}
Note that a finite mixing $v\neq 0$ occurs only due to the tetrahedral CEF contribution ($y\neq 0$) which leads to $d\neq 0$ in Eq.~(\ref{1.02}) and hence $\delta\neq 0$. When $d=0$ $(\delta = 0)$ there is no mixing between $|0,0\rangle$ and  $|1,0\rangle$ and consequently the energies of  $|\Gamma_s\rangle$ and $|\Gamma_t^0\rangle$ will be independent of the field H. For nonzero $d$ and $v$ these two levels will repel with increasing field H . The other two triplet states  $|\Gamma_t^\pm\rangle$ have a linear Zeeman splitting of 2h independent of d. For $\delta\geq 0$  the singlet ground state level $E_s$ and lowest triplet level $E_t^+$ cross at a critical field $h_c=\Delta/(1-\delta^2)$ meaning $\delta_+(h_c)=0$.

\begin{table}[t]
\caption{Singlet-triplet CEF states, levels and excitation energies in a magnetic field ${\bf H} \parallel [001]$. Here $\delta_n(H)=E_t^n(H)-E_g(H)$. The eigenstates are given in terms of zero-field singlet-triplet states $|0,0\rangle$ and $|1,\pm\rangle, |1,0\rangle$ respectively ($h=g\mu_BH$). }
\begin{center}
\begin{tabular}{cccc}
%\hline
eigenstate  & $|\Gamma(H)\rangle$ $\qquad $  & E(H) $\qquad$ & $\delta_n(H)$ $\qquad$\\
\hline
$|\Gamma_s(H)\rangle$ & $u|0,0\rangle + v|1,0\rangle$ &$\frac{1}{2}(\Delta-\tilde{\Delta})$ & 0 \\
\hline
$|\Gamma_t^+(H)\rangle$ &$|1,+\rangle$  & $\Delta - h$ & $\frac{1}{2}(\Delta+\tilde{\Delta})-h$ \\
%\hline
$|\Gamma_t^0(H)\rangle$ &$u|1,0\rangle - v|0,0\rangle$ &$\frac{1}{2}(\Delta+\tilde{\Delta}$) & $\tilde{\Delta}$\\
%\hline
$|\Gamma_t^-(H)\rangle$ & $|1,-\rangle$ & $\Delta + h$ & $ \frac{1}{2}(\Delta + \tilde{\Delta})+h$\\
\hline
\end{tabular}
\end{center}
\label{table:table1}
%\centerline{\epsfxsize=6.0truecm \epsfbox{fig1.eps} }
\end{table}

For evaluation of the effective mass we need the quadrupolar matrix elements in Eq.~(\ref{2.19}). They may all be expressed in terms of the irreducible zero field matrix elements  $\alpha' =\frac{\sqrt{3}}{4}(13-20d^2)$
and $\beta'=\sqrt{35(1-d^2)}$. With their help and defining $\delta'=\alpha'/\beta'$ one obtains the following nonzero matrix elements:

%23
\begin{eqnarray}
|\langle\Gamma_s|O_{yz}|\Gamma_t^+\rangle|^2 = |\langle\Gamma_s|O_{zx}||\Gamma_t^-\rangle|^2
&=& \frac{1}{2}\beta'^2(u-\delta' v)^2=|m_Q^-|^2\nonumber\\
|\langle\Gamma_s|O_{yz}|\Gamma_t^-\rangle|^2 = |\langle\Gamma_s|O_{zx}||\Gamma_t^+\rangle|^2
&=& \frac{1}{2}\beta'^2(u+\delta' v)^2=|m_Q^+|^2\\
|\langle\Gamma_s|O_{xy}|\Gamma_t^0\rangle|^2
&=&\beta'^2 = |m^0_Q|^2\nonumber
\label{2.23}
\end{eqnarray}

Inserting the matrix elements and excitation energies in Eq.~(\ref{2.19}) and using
 $|m_Q^+|^2+|m_Q^-|^2| =\beta'^2[1-v^2(1-\delta'^2)]$ we finally obtain the expression

%24
\begin{equation}
\chi_Q(H)=\frac{2\beta'^2}{\Delta}\Bigl[\frac{2\Delta\Delta'}{\Delta'^2-h^2}[1-v^2(1-\delta'^2)]
+\frac{\Delta}{\tilde{\Delta}}\Bigr]
\label{2.24}
\end{equation}

where we defined $\Delta'=\frac{1}{2}(\Delta+\tilde{\Delta})$. Without tetragonal CEF ($d^2=0$) we have $\Delta' = \tilde{\Delta}=\Delta$  and then the above expression reduces to

%24a
\begin{equation}
\chi_Q(H)=\frac{2\beta'^2}{\Delta}\Bigl[\frac{2\Delta^2}{\Delta^2-h^2}[1-v^2(1-\delta'^2)] + 1\Bigr]
\label{2.24a}
\end{equation}

For small fields ($h\ll\Delta$) the general $\chi_Q(H)$ in Eq.~(\ref{2.24a}) may be expanded with a leading term $\sim (h/\Delta)^2$ according to

%25
\begin{equation}
\chi_Q(H)\simeq \frac{2\beta'^2}{\Delta}
\Bigl\{3+\bigl[2\delta^2(\delta'^2-3)\bigr] \Bigl(\frac{h}{\Delta}\Bigr)^2\Bigr\}
\label{2.25}
\end{equation}

The zero field mass enhancement without tetragonal CEF ( $d^2=0$) is then obtained from Eq.~(\ref{2.19},\ref{2.24a}) simply as

%26
\begin{equation}
\frac{m^*}{m} = 1 + g^2 N(0)3\bar{f}\frac{2\beta'^2}{\Delta}
\label{2.26}
\end{equation}

The states and energies in Table~\ref{table:table1} are nominally derived for ${\bf H}\parallel [001]$. However it was shown in Refs.~\onlinecite{Shiina04a,Shiina04b} that for low fields $(h<\Delta)$ they are the same for all field directions, i.e. approximately  isotropic. Therefore the quadrupolar susceptibility derived above and the related mass enhancement are also approximately isotropic as long as h is appreciably below $\Delta$.  This condition is required anyway in the dispersionless case where the calculation is only valid for moderate fields when $\delta_+(h)$ is still large enough.

%
%%%%%%%%%%%%%%%%%%%%%%%%%%%%%%%%%%%%%%%%%%%%%%%%%%%%%%%%%%%%%
\begin{figure}
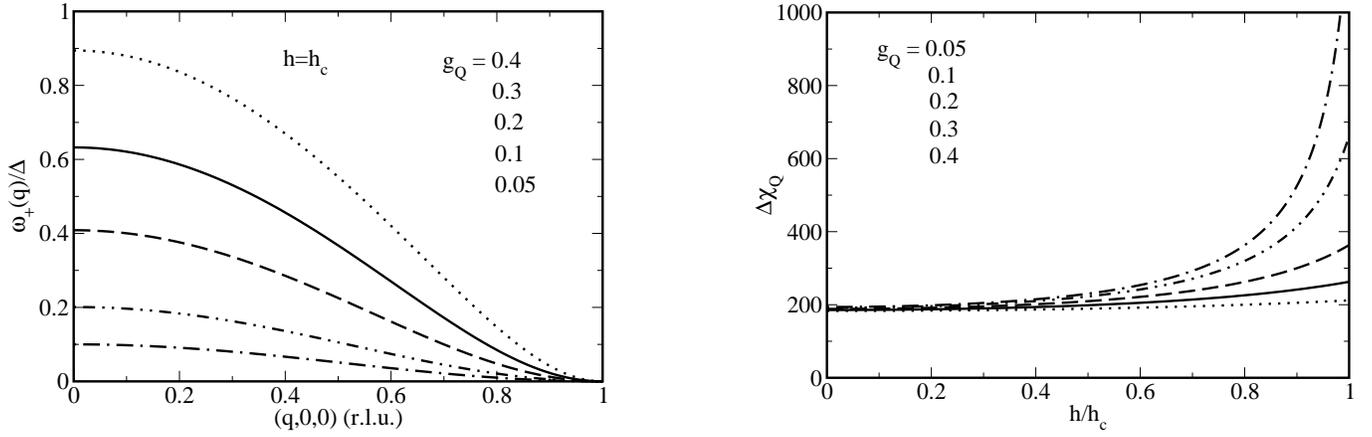

\includegraphics[width=8.0cm]{Fig3a}\hfill
\includegraphics[width=8.0cm]{Fig3b}
\caption{Left (a): Dispersion of lowest triplet quadrupolar exciton  $\omega_+(\v q)$  at the critical field $h_c(g_Q)$ where
 $\omega_+(\v Q)$ becomes soft (Q in units of $2\pi/a$). As g$_Q$ is reduced the dispersion becomes flat increasing the
 phase space for low energy conduction electron scattering. Right (b): Field dependence of $\Delta\chi_Q\sim \delta m^*/m$ for various strengths of intersite quadrupolar coupling $g_Q$ (left). For small $g_Q$ mass renormalisation close to $h_c$ is large due to flat $\omega_+(\v q)$  dispersion. For larger $g_Q$ the dispersion becomes stronger and $\omega_+(\v q)$ softens only in the vicinity of $\v Q = (0,0,1)$ leading to a much smaller mass enhancement at $h_c$. The curves correspond to $g_Q$ given in the legend in decreasing order. For the value $g_Q = 0.3$ corresponding to \POS~ little field dependence of $\delta m^*/m$  remains.}
\label{fig:Fig2}
\end{figure}
%%%%%%%%%%%%%%%%%%%%%%%%%%%%%%%%%%%%%%%%%%%%%%%%%%%%%%%%%%%%%
%%%%%%%%%%%%%%%%%%%%%%%%%%%%%%%%%%%%%%%%%%%%%%%%%%%%%%%%%%%%%
\begin{figure}
\raisebox{-0.3cm}
{\includegraphics[width=9.2cm]{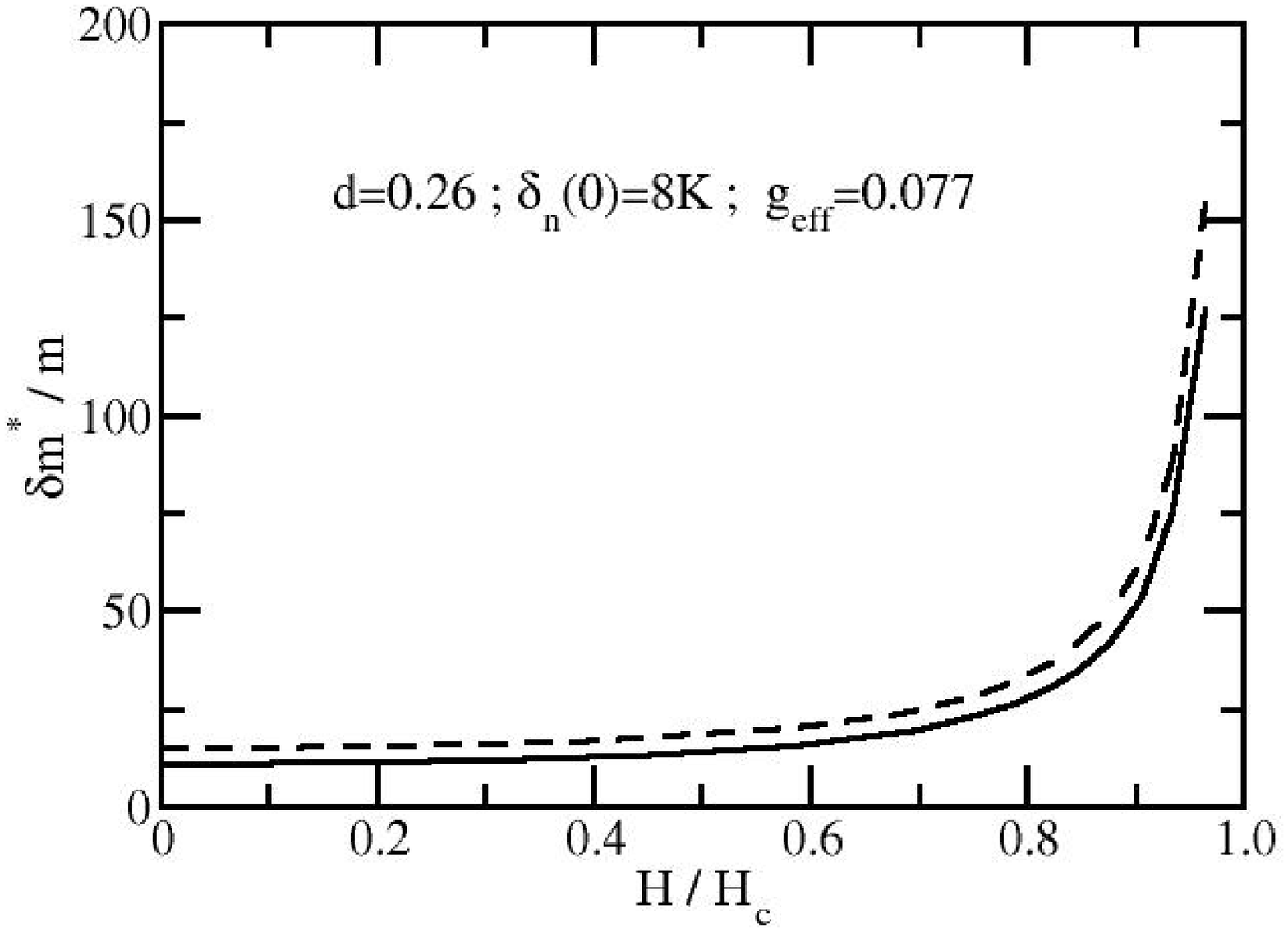}}\hfill
\includegraphics[width=8.0cm]{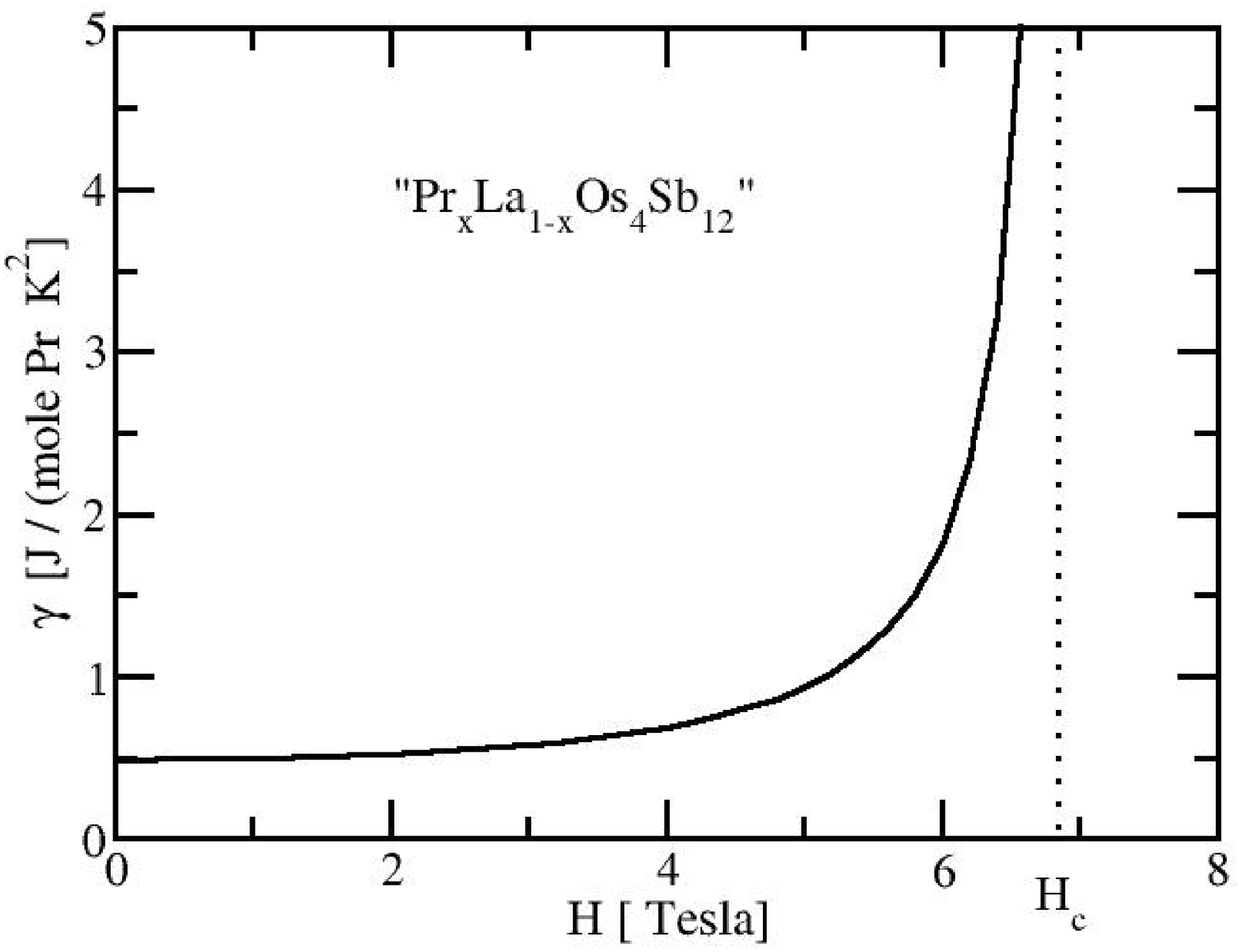}
\caption{Left (a): Variation with magnetic field of $\delta m^*/m$ calculated from the self consistent calculation for finite band width
$\epsilon_c =20 \delta_n (0)$ (full line) and $\epsilon_c \to \infty$ (broken line). Right (b): Variation with magnetic field of the specific heat coefficient $\gamma$ 
calculated selfconsistently for $\epsilon_c =20 \delta_n (0)$}
\label{fig:Fig22}
\end{figure}
%%%%%%%%%%%%%%%%%%%%%%%%%%%%%%%%%%%%%%%%%%%%%%%%%%%%%%%%%%%%%

\section{Influence of quadrupole exciton dispersion on the mass enhancement}

\label{sect:disp}

In the previous section we investigated a model of noninteracting local singlet-triplet quadrupole excitations. For appreciably large concentration of Pr ions in the system \POSX~ this is no longer justified. Due to effective interactions between the 4f states on different sites the singlet-triplet excitations at $\Delta$ acquire a dispersion. Formally this is already included in the self energy of Eq.~(\ref{2.15}) provided the boson propagator for a dispersive mode is used by replacing $\delta_n\rightarrow \omega_n(\v q)$ in Eq.~(\ref{2.17}). The dispersion is due to quadrupolar RKKY-type intersite interactions which are obtained in second order perturbation theory from H$_{AC}$ and given by \cite{Shiina04c}

%28
\begin{equation}
H_Q=\sum_{\langle ij\rangle}K_Q(ij)\v O(i)\cdot \v O(j)
\label{2.28}
\end{equation}

where $\v O =(O_{yz},O_{zx},O_{xy})$ is the $\Gamma_5$ type quadrupole. The sum is restricted to nearest neighbors and $K_Q$ is the effective quadrupolar coupling constant. This interaction leads to the field induced antiferroquadrupolar order from which the $\Gamma_5$ symmetry has been infered \cite{Shiina04a}. Formally H$_Q$ may be obtained from an RKKY type mechanism in order $\sim g^2$ in the coupling constant of H$_{AC}$.
%29
%\begin{equation}
%K_Q(ij)=g^2\sum_{\v k,\v q} \hat{q}^2_\alpha\hat{q}^2_\beta
%\frac{f_{\v k + \v q } - f_{\v k}}{\epsilon_{\v k}-\epsilon_{\v k + \v q }}\exp(i\v q\cdot(\v R_i-\v R_j))
%\label{2.29}
%\end{equation}
In practice the n.n. term $K_Q$ is determined from the experimentally observed dispersion of the quadrupolar excitons  $\omega_n(\v q)$ which are fully degenerate for zero field \cite{Kuwahara05}. In finite field the dispersive excitation branches are obtained by replacing the Hamiltonian in
Eq.~(\ref{2.09}) with $H_{CEF}+H_Q+H_Z$. Using  a generalised Holstein-Primakoff approximation \cite{Shiina04c} the three quadrupolar exciton modes at moderate fields are described by

%30
\begin{eqnarray}
\omega_\pm(\v q)&=&\omega({\v q})\mp h; \quad \omega_0(\v q)=\omega(\v q)
\label{2.30}
\end{eqnarray}

where the zero-field dispersion is given by

%31
\begin{eqnarray}
\omega(\v q)&=&\sqrt{A^2_{\v q}-B^2_{\v q}} \nn\\
A_{\v q} &=& \Delta+zK_Q\gamma_{\v q}; \quad B_{\v q} = -zK_Q\gamma_{\v q}\\
\gamma_{\v q}&=&\cos\frac{1}{2}q_x\cos\frac{1}{2}q_y\cos\frac{1}{2}q_z
\label{2.31}
\end{eqnarray}

Here z=8 is the coordination number and $\gamma_{\v q}$ the structure function of the bcc cubic lattice of Pr ions with
momentum {\v q} measured in r.l.u. $(2\pi/a)$. The width of the exciton dispersion is controled by the effective quadrupolar
coupling constant $K_Q$ or in dimensionless form by $g_Q=z{\beta'}^2K_Q/\Delta$. From the analysis of the AFQ phase diagram \cite{Shiina04c}  and  experimental zero-field dispersion \cite{Kuwahara05} one may deduce $g_Q\simeq 0.3$ in \POS. The minimum of the dispersion occurs at the bcc zone boundary wave vector \v Q = (1,0,0) (r.l.u.). The zero-field energy is given by $\omega(\v Q)=\Delta[1-2g_Q]^\frac{1}{2}$. Consequently the soft mode indicating transition to (zero-field) AFQ order would occur at $g_Q=0.5$ which is larger than the above value of 0.3 for pure \POS. Therefore application of a magnetic field is necessary to achieve a soft mode $\omega_+(\v Q) = 0$ at a critical field $h_c$. The  dispersions in Eq.~(\ref{2.31}) are approximations where the field dependence of the $\Gamma_s -\Gamma_t^0$ splitting has  been neglected. This is possible as long as their  dipolar matrix element $d^2, \delta^2 \ll 1$ which is true for the case $d^2 = 0.067$ (Fig.~\ref{fig:Fig1} (left))  Then the soft mode condition leads to the approximate critical field $h_c/\Delta =(1-2g_Q)^\frac{1}{2}$ above which AFQ order will be induced. Using $g_Q = 0.3$ leads to $h_c/\Delta =0.586$ which is close to the exact value 0.632 given in Ref.~\onlinecite{Shiina04c}.\\

Calculation of electron self energy and mass enhancement in the dispersive case proceeds now exactly along the lines described in Sect.~\ref{sect:mass}. The main modifications arise from the fact that a more sophisticated approximation for the electron-quadrupolar exciton spectral function is employed. For large cut-off energies $\epsilon_c \to \infty$ one obtains

%32
\begin{eqnarray}
\frac{m^*}{m} &=& 1 + g^2 N(0) \sum_{\alpha\beta,n}
2|\langle \Gamma_s \left| O_{\alpha\beta} \right| \Gamma^n_t\rangle|^2
\frac{1}{2\pi}\int d\Omega_{\v q}
\frac{\hat{q}^2_\alpha \hat{q}^2_\beta}{\omega_n(\v q)}
\label{2.32}
\end{eqnarray}

which closely parallels the expression derived by Nakajima and Watabe \cite{Nakajima63} for the effective mass enhancement due to electron-phonon interaction.
Here we use $\v q = q\hat{\v q}$ with $q=2p_F\cos\theta$  (Fig.~\ref{fig:Fig0}) where $\hat{q}$ has polar and azimuthal angles $\theta$ and  $\phi $, respectively. Due to the geometric restrictions only half the solid angle ($2\pi$) contributes in the momentum integral.
Replacing the dispersive modes $\omega_n(\v q)$ by the dispersionless singlet-triplet excitation energies $\delta_n$ leads to the previous result in Eqs.~(\ref{2.19},\ref{2.24a}). Using the explicit matrix elements and dispersions we may again represent  the mass enhancement in the form of Eq.~(\ref{2.19})

%33
\begin{eqnarray}
\frac{m^*}{m}& =& 1 + g^2 N(0)\bar{f}\chi_Q(H)\nn\\
\chi_Q(H)&=&\frac{2\beta'^2}{\Delta}\frac{1}{\bar{f}} \Bigl\{[1-v^2(1-\delta'^2)]
\frac{1}{2\pi}\int d\Omega_{\v q}\frac{\Delta\omega(\v q)}{\omega(\v q)^2-h^2}
\hat{q}^2_z(\hat{q}^2_x+\hat{q}^2_y)+
\frac{1}{2\pi}\int d\Omega_{\v q}\frac{\Delta}{\omega(\v q)}
\hat{q}^2_x\hat{q}^2_y \Bigr\}
\label{2.33}
\end{eqnarray}

Here the first and second terms are due to the virtual excitations of $\omega_\pm({\bf q})$ and  $\omega_0({\bf q})$ bosons respectively.
In the dispersionless limit this reduces to the previous result in Eq.~(\ref{2.24a}) for the case $d^2=0$ (no tetragonal CEF) which corresponds to the present treatment due to the neglect of the  $E_s,E_t^0$ level repulsion implied in the dispersions of Eq.~(\ref{2.30}). The above expression for the mass enhancement have to be evaluated numerically due to the BZ integrations. This will be discussed in the next section. But we may nevertheless gain some qualitative insights by simple approximations to these integrals in the zero-field case. For that purpose we expand $\omega(\v q)$ around one of the six equivalent zone boundary X- points with \v Q =($\pm1,0,0$) {\em etc.}. Then one obtains an isotropic approximate dispersion given by

%34
\begin{eqnarray}
\omega(\v q')^2=\omega_{\v Q}^2+\omega_0^2(\pi q')^2
\label{2.34}
\end{eqnarray}

where $ \v q'=\v Q- \v q$ is the momentum vector counted from the X-point  and q' is its length. Furthermore $\omega _{\v Q} = \Delta[1-2g_Q]^\frac{1}{2} = h_c$ and $\omega_0 = \sqrt{g_Q}\Delta$. On approaching the critical $g_Q^c=\frac{1}{2}$ the soft mode frequency $\omega_{\v Q}$ vanishes. In this limit the integral in Eq.~(\ref{2.32}) may easily be evaluated.  The approximate Fermi surface geometry of \POS~ shows that $p_F\simeq\frac{1}{\sqrt{2}}$ (r.l.u.) or $2p_F=\sqrt{2}$ (Fig.~\ref{fig:Fig0}). Therefore $2p_F > Q$ which means that the minimum in $\omega(\v q)$ is included in the domain of the momentum integrals in Eqs.~(\ref{2.32},\ref{2.33}). We restrict the latter to the sphere around the minimum at \v Q (X-point) with a cutoff radius given by $q'_c < 1$ where only 1/2 contributes due to geometric restrictions. We then obtain from Eq.~\ref{2.32}

%26
\begin{equation}
\frac{m^*}{m} \simeq 1 + g^2 N(0)\bigl(\frac{{q'}_c^4}{4\pi}\bigr)\frac{2\beta'^2}{\omega_0}
\label{2.35}
\end{equation}
where the momentum cutoff  $q'_c$  around X is defined such that the quadratic expansion in  Eq.~(\ref{2.34}) is still valid.
Note that although the exciton energy becomes soft $\omega_{\v Q}\rightarrow 0$ at the zone boundary there is no divergence in the mass renormalisation. This is due to the small phase space volume around the X-point which gives only a small contribution despite the vanishing exciton frequency. In addition the singular contribution is suppressed by the fact that the form factor $\hat{q}^2_\alpha \hat{q}^2_\beta $  vanishes exactly at the X-point directions and only contributions from its environment are picked up by the integration. The above expression is  formally quite similar to the dispersionless result  of Eq.~(\ref{2.26}). In the latter case the renormalisation diverges when $\Delta\rightarrow 0$ because this corresponds to a softening in the whole BZ. Thus we conclude from Eqs.~(\ref{2.26},\ref{2.35}) that the inclusion of a mode dispersion removes the problem of singular mass renormalisation, $m^*/m$ stays finite for all coupling constants $g_Q$, even when $g_Q=g_Q^c$ when the exciton frequency becomes soft at \v Q.
However the above formula cannot give a reliable estimate for $m^*/m$ due to the strong momentum cutoff dependence, we therefore have to employ a numerical evaluation of Eq.~(\ref{2.33}).

%%%%%%%%%%%%%%%%%%%%%%%%%%%%%%%%%%%%%%%%%%%%%%%%%%%%%%%%%%%%%
\begin{figure}
\includegraphics[width=8.0cm]{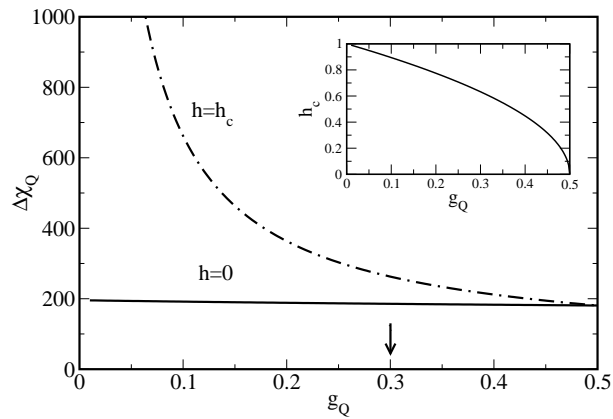}
\caption{Quadruplar susceptibility  $\Delta\chi_Q\sim \delta m^*/m$ as function of effective intersite coupling $g_Q$.
For zero field the intersite coupling or dispersive width of $\omega (\v q)$ has little effect on the  mass enhancement. It actually decreases slightly when $g_Q^c = 0.5$ is approached despite the appearance of the soft mode $\omega(\v Q)\rightarrow 0$. At the critical field h$_c$ the dispersion has much stronger influence:
when the latter becomes small (decreasing $g_Q$) the mass enhancement at $h_c$ strongly increases. For \POS~ ($g_Q=0.3$) one may expect little field dependence of $\delta m^*/m$ between  $h=0$ and $h=h_c$. The inset shows the dependence of the AFQ critical field $h_c$ on $g_Q$ with the approximation $d^2\simeq 0$.}
\label{fig:Fig3}
\end{figure}
%%%%%%%%%%%%%%%%%%%%%%%%%%%%%%%%%%%%%%%%%%%%%%%%%%%%%%%%%%%%%

\section{Numerical results and discussion}

\label{sect:discussion}

We first discuss the mass enhancement in the  large band width approximation $\epsilon_c \to \infty$ for dispersionless undamped CEF excitations.
The absolute value of $m^*(h)/m=1+\delta m^*/m$  is determined by the effective coupling constant $\tilde{g}=g^2N(0)$. Approximating  $N(0)\sim1/2W$ (2W=band width) this may be written as $\tilde{g}/\Delta=\lambda^2(W/\Delta)$. Here we introduced the dimensionless quadrupolar coupling constant $\lambda =gN(0)$. Assuming typical values of W =1 eV, $\lambda\simeq 0.02$ and using $\Delta$ = 8 K we obtain $\tilde{g}/\Delta\bar{f}\simeq 0.077 $ as the size of the effective coupling for the quadrupolar mass enhancement mechanism. Using $d^2=0.067$ and hence $\beta'^2\simeq 32.6$ we obtain a zero field enhancement of $m^*/m\simeq 16$. This corresponds to the right magnitude for thermal mass enhancement.\\

For discussion of the field dependence we use the quadrupolar susceptibility which contains only $d$ as adjustable parameter to avoid specifying $\tilde{g}$. In the case of weak tetrahedral CEF such as realised in \POSX~ the level repulsion of $\Gamma_s$ and $\Gamma^0_t$ is also weak and therefore the $\Gamma^+_t$ level crosses the $\Gamma_s$ ground state at a critical field $h_c = \Delta/(1-\delta^2)$ in the dispersionless case. The decrease in the excitation gap for $h<h_c$ and the field dependence of matrix elements leads to a field dependence of $\chi_Q(h)$ (Eq.~(\ref{2.24})) which is shown in Fig.~\ref{fig:Fig1} for $d^2=0$ and $d^2=0.067$.
For larger $d^2$ the increase is diminished and eventuallly for $d^2>0.42$ the level repulsion due to the tetrahedral CEF is strong enough to lead to an {\em increase} in excitation energy and hence to a {\it decreasing} effective mass. Likewise the level repulsion prevents field induced AFQ order, therefore it is not appropriate for concentrated \POS. This case resembles more that of pure metallic Pr where a singlet-singlet level repulsion in a field also leads to a decrease in quasiparticle mass \cite{Fulde83}.\\

Let us next turn to the divergent mass renormalization which is predicted for dispersionless undamped CEF excitations  when the triplet level $E_t^+$ approaches the singlet ground state level $E_s$ (Fig.~\ref{fig:Fig1}). Therefore in the (level crossing) case $d^2<0.42$ this approach  is only valid for moderate fields.
As the divergence follows directly from the general analytic structure of the corresponding electron self energy it persists also in the self consistent solution for finite band width. Selfconsistency leads to an overall reduction of the mass renormalization as can be seen from Fig.~\ref{fig:Fig22}.

The (unphysical) divergence of the mass enhancement close to the critical magnetic field is an artefact of the model which assumes dispersionless undamped CEF excitations. Inelastic neutron scattering \cite{Kuwahara05}
however, have shown that the singlet-triplet excitations have a pronounced dispersion, The dispersive width corresponds to $\sim 40\%$ of the CEF excitation energy $\Delta$. An applied field of critical strength therefore leads to a softening of $\omega(\v q)$ only in the restricted phase space around the AFQ ordering vector \v Q. Consequently the mass renormalisation will be finite even at the critical field $h_c$ for AFQ order when  $\omega(\v Q)$=0. This is shown in Fig.~\ref{fig:Fig2}a  for various effective quadrupolar coupling strenghts $g_Q$. For  $g_Q=0$ the mass renormalisation at $h_c$ would diverge as in Fig.~\ref{fig:Fig1} because the excitation energy becomes soft for all \v q-vectors. For small g$_Q$ and dispersive width the softening apperars only around \v Q (Fig.~\ref{fig:Fig2}b) and the mass enhancement is finite, though still large at $h_c$. It is progressively diminished with further increasing $g_Q$ because the softening becomes strongly constricted around \v Q (see Fig.~\ref{fig:Fig2}b). For the value $g_Q =0.3$  appropriate for \POS~ the dispersion is sufficiently large to suppress the field dependence of $\delta m^*/m$ as depicted in Fig.~\ref{fig:Fig2}a  (full line).\\

An alternative presentation of results for the dispersive case is shown in Fig.~\ref{fig:Fig3}. The mass enhancement is shown as function of quadrupolar coupling $g_Q$ ($\sim$ dispersive width)  for the two limiting cases $h=0$ and $h=h_c(g_Q)$. For large $g_Q$ and dispersion the field variation of $\delta m^*/m$  between $h=0$ and $h=h_c$ becomes small. This is partly due to the fact that $h_c(g_Q)$ itself becomes small for large $g_Q$ (see inset of Fig.~\ref{fig:Fig3}). When g$_Q$ decreases the difference in $\delta m^*/m$ for $h=0,h_c$ increases rapidly because the mass enhancement at $h=h_c$ becomes singular when approaching the dispersionless case $g_Q\rightarrow 0$. The arrow corresponds to the proper value of $g_Q$ for \POS~ and it shows again that one should expect little field dependence of the mass enhancement in this case.

\section{Conclusion and outlook}

\label{sect:conclusion}

In this work we have studied in detail the quasiparticle mass enhancement originating in the aspherical Coulomb scattering of conduction electrons from singlet triplet CEF excitations. This model has some relevance for the heavy fermion superconductor \POS~ where the Pr$^{3+}$  4f states are subject to a tetrahedral CEF leading to a singlet  ground state and an excited triplet. For small tetrahedral CEF the latter has a mostly nonmagnetic character and therefore may be excited by aspherical Coulomb scattering from conduction electrons. These virtual second order processes lead to a quasiparticle mass renormalisation which may well be the source of the large thermal  and dHvA effective masses observed in \POS. A hybridsation mechanism between conduction and 4f electrons  can be ruled out since the Fermi surface of \POS~ is identical to that of \LOS~ which advocates for fully locallised 4f electrons in Pr. Indeed the well defined CEF excitations seen in INS \cite{Kuwahara05} support this view.\\

If aspherical Coulomb scattering of conduction electrons plays a role in the mass enhancement one should expect a field dependence of the latter because the triplet excited state splits in the field. For small enough tetrahedral CEF characterised by the parameter $d^2\ll 1$ the lowest triplet component crosses the singlet ground state at a critical field $h_c$. The mass enhancement in second order perturbation theory with respect to $H_{AC}$ then increases with field and becomes singular at $h_c$. For larger tetrahedral CEF (d$^2>0.42$) the excitation energy between singlet ground state increases with field leading to a decrease of the mass enhancement, similar as has been observed in Pr metal where the mass renormalisation is due to exchange scattering from a singlet-singlet CEF level scheme.\\

The observed singular mass enhancement close to the critical field of level crossing is an artefact of the dispersionless model, both in the perturbative and selfconsistent  treatment. Any dispersion of the singlet-triplet excitations due to effective intersite quadrupolar interactions will lead to a finite effective quasiparticle mass. We have shown that the enhancement decreases strongly with increasing dispersion because the phase space for conduction electron scattering from low lying CEF excitations (quadrupolar excitons) is constrained to the wave vector \v Q of incipient field induced AFQ order. For a quadrupolar coupling constant $g_Q=0.3$ corresponding to \POS~ the field dependence is reduced to a few percent. In addition this compound is superconducting below T$_c$ = 1.85 K with $H_{c2}$ = 2.2 T and has a field induced AFQ phase above $H_c$ = 4.5 T. Therefore only a reduced field range is left to observe the small field dependence possible at $g_Q=0.3$. We conclude that the concentrated \POS~ is not a favorable system to observe the field dependent mass enhancement due to aspherical Coulomb scattering.\\

A more promising system may be the La-diluted systems \POSX. On increasing x the average distance between the Pr 4f shells becomes larger and therefore the effective quadruplar coupling $g_Q(x)$ will decrease with x, i.e. the dispersion of the 4f CEF ecxitons will progressively decrease. This means that the field dependence of effective masses will become more pronounced according to Figs. \ref{fig:Fig2},\ref{fig:Fig3}. Of course the {\em absolute} (zero-field) size of the mass enhancement is also reduced since the self energy in Eq.~(\ref{2.18a}) will be proportional to the number (1-x) of Pr sites. There should however be an intermediate concentration region for x where the field dependence is pronounced (g$_Q$ small) and the $\gamma(x)$ still large enough as compared to the other (lattice or CEF-Schottky) contributions such that the field dependence of $\gamma(x,H)$ is experimentally accessible. Furthermore in this region of x one may probe a larger field range because there is no more  AFQ order present.  Therefore we propose that the field dependence of the electronic specific heat in mixed crystals of \POSX~ is systematically investigated and analysed. It may hold important clues to the microscopic origin of the large effective mass in the concentrated compound \POS.\\[2cm]
{\it Acknowledgement}\\
The authors would like to thank R. Shiina for useful comments.

%%%%%%%%%%%%%%%%%%%%%%%%%%%%%%%%% %%%%%%%%%%%%%%%%%%%%

\end{document}